\documentclass[pra,showpacs,superscriptaddress,twocolumn]{revtex4}
\usepackage{graphicx}
\usepackage{subfigure}

\begin{document}

\title{Thermodynamic witness of quantum probing}

\author{H.Dong}
\affiliation{Institute of Theoretical Physics,Chinese Academy of
Science, Beijing 100080, China}

\author{X.F. Liu}
\affiliation{Department of Mathematics, Peking University, Beijing 100871, China}

\author{C.P. Sun}
\email{suncp@itp.ac.cn}
\homepage{http://power.itp.ac.cn/~suncp/}
\affiliation{Institute of Theoretical Physics,Chinese Academy of Science, Beijing 100080,
China}

\begin{abstract}
The thermodynamic influence of quantum probing on an object is
studied. Here, quantum probing is understood as a pre-measurement
based on a non-demolition interaction, which records some
information of the probed object, but does not change its energy
state when both the probing apparatus and the probed object are
isolated from the environment. It is argued that when the probing
apparatus and the probed object are immersed in a same equilibrium
environment, the probing can affect the effective temperature of the
object or induce a quantum isothermal process for the object to
transfer its energy. This thermodynamic feature can be regarded as a
witness of the existence of quantum probing even if the quantum
probing would not disturb the object if the environment were not
present.
\end{abstract}

\maketitle

\narrowtext

\textit{Introduction}- The Landauer's principle that the erasure of one bit
of information requires a minimum heat generation of $k_{B}T\ln2$, which is
based on the second law of thermodynamics \cite{landaure,book_maxwelldeamon2}%
, underlies the thermodynamics of information processing. This
principle eventually resolves the Maxwell's demon paradox: why a
demon can assist a binary thermal medium to do extra work
\cite{book_maxwelldeamon2}. For the science and technology of
quantum information, the Landauer's principle is undoubtedly crucial
since it gives a physical limitation on the spatial-time scales of
logical devices on chips.

However, in the arguments \textit{for} and \textit{against} the
Landauer's principle, the conventional question in the
thermodynamics of information processing that whether the
measurement process requires a cost of heat generation has never
been convincingly answered. Surely, earlier authors had touched on
this problem, but it has not been clarified yet because the
measurement process is not defined properly. Particularly, in the
quantum  approach of measurement, people can not be unanimous for
some fundamental problems, such as whether or not there exist wave
function collapse \cite{zurek}. In fact, to clarify the situation,
we need to answer the following subtle questions in an unambiguous
way: when can we say a system is performing a measurement on another
system and what kind of measurement can dissipate information?

In this paper, we refer quantum probing to a pre-measurement
\cite{book_measurement,sun} based on the non-demolition coupling of
the probed system $S$ to the apparatus $A$, which is only a unitary
process to produce entanglement between $S$ and $A$, and does not
concerns the subtle, seemingly philosophical arguments, such as wave
function collapses. Generally speaking, when the energy state of the
measured system $S$ is not influenced by the coupling to the
measuring apparatus $A$, but $A$ can record some information of the
system $S$, the measurement performed by $A$ on $S$ is called a
pre-measurement. The so called non-demolition (pre-) measurement
\cite{book_measurement} is an ideal measurement under some
circumstances. In the case of non-demolition measurement, the
system-apparatus coupling $V_{SA}$ commutes with the Hamiltonian
$H_{S}$ of the measured system, but does not commute with the
Hamiltonian $H_{A}$ of the measuring apparatus. When studying
non-demolition measurement it is usually assumed that both $A$ and
$S$ are isolated from an environment. In this paper, we will
investigate the effect of environment on the measured system in a
non-demolition measurement. In the following discussion, we will use
the name quantum probing or just probing when regarding such
non-demolition measurement. In the presence of environment, the
practical thermalization \cite{efftem} of the total system $A+S$
will have thermodynamic effects on $S$. These effects can be
regarded as the thermodynamic witness of quantum probing.

\begin{figure}[tpb]
\includegraphics[width=7cm,bb=85 485 536 655]{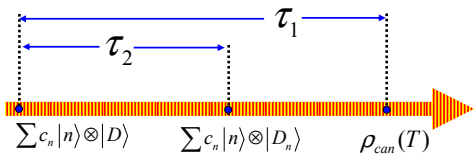}
\caption{Time scale of the non-demolition pre-measurement and
thermalization. (1) In the time interval $\left(0,\protect\tau_{2}\right)$,
the apparatus makes the pre-measurement: $\sum c_{n}\left|n\right\rangle
\otimes\left|D\right\rangle \rightarrow\sum c_{n}\left|n\right\rangle
\otimes\left|D_{n}\right\rangle $ and the environment does not play the
role, since its coupling to $S+A$ is weaker than that between $S$ and $A$.
(2) In the time interval $\left(\protect\tau_{2},\protect\tau_{1}\right)$,
the effect of thermalization due to the environment becomes prominent. The
total system $S+A$ is finally thermalized in the canonical state $\protect%
\rho_{\mathrm{can}}\left(T\right)$ with temperature $T$.}
\label{fig:timescale}
\end{figure}

When the existence of an environment $E$ is considered, quantum probing is
understood in two steps(illustrated in Fig. \ref{fig:timescale}) :

\begin{itemize}
\item Due to the extreme-weakness of the coupling between $S+A$ and $E$, it
can be neglected within the dephasing time $\tau_{2}$ of the system, the
dephasing being the result of the interaction between $S$ and $A$. In the
end of this step, the initial factorized state of $S+A$ becomes a state
assuming the form of an ideal Schmidt decomposition $\sum
c_{n}\left|n\right\rangle \otimes\left|D_{n}\right\rangle $\cite{sun}, with $%
\left\langle D_{m}\right|\left.D_{n}\right\rangle =\delta_{mn}$.

\item If the probing apparatus continues to probe the state of the system $S$%
, the non-demolition coupling should hold for a longer time, and the
communication with the environment will result in the thermalization of the
total system $S+A$. In this step, the information of the initial state
should be erased totally, but the correlation between $A$ and $S$ needs to
exist to leave the witness of quantum probing.
\end{itemize}

As shown as follows, it is the comment environment of the probed
system and the probing apparatus that selects a special set of ideal
entanglement states of $S+A$ to be thermalized, so that these
witnesses of quantum probing are of thermodynamics, and thus
observable in the classical or macroscopic level.

\textit{Universal Setup for quantum probing apparatus}- Let $A$ be a general
apparatus weakly coupling to the system $S$ to be probed. We require that
the energy spectrum of $A$ be denser than that of $S$. The following
heuristic argument may help to justify this requirement: to measure the
spatial scale of an object, the ruler should have a much finer graduation
than the size of this object.

To be precise, let $H_{A}=\sum_{k}\epsilon_{k}\left|k\right\rangle
\left\langle k\right|$ be the spectrum decomposition of the Hamiltonian of $%
A $ and $H_{S}=\sum_{n}E_{n}\left|n\right\rangle \left\langle n\right|$ the
spectrum decomposition of the Hamiltonian of $S$, the requirement then can
be expressed as $\min\{|E_{n}-E_{n+M}|\}>>\max\{|\epsilon_{k}-\epsilon_{k+1}|\}$. Here, $%
\left|k\right\rangle$ is the eigenvector of $H_{A}$ corresponding to the
eigenvalue $\epsilon_{k}$, and $\left|n\right\rangle$ is the eigenvector of $%
H_{S}$ corresponding to the eigenvalue $E_{n}$ Let $V_{AS}$ be a
weak coupling between $S$ and $A$. The weakness of $V_{AS}$ means
that its effect on the dynamics of the total system $S+A$ can be
well studied by the perturbation method \cite{frohlich}.

To investigate the behavior of the total system $S+A$ immersed in an
environment, we consider the partition function
\[
Z=\mathrm{Tr}\left( e^{-\beta H}\right) =\mathrm{Tr}\left( W^{\dagger
}e^{-\beta H}W\right)
\]%
Here, as a trick, we have introduced a unitary transformation $W=\exp \left(
-S\right) $, defined by an anti-Hermitian operator $S$, which is a
perturbation quantity of the same order as $V_{AS}$. If
\[
V_{AS}+\left[ H_{A}+H_{S},S\right] =0,
\]%
then the partition function can be approximated as $Z=\mathrm{Tr}\exp \left(
-\beta H_{\mathrm{eff}}\right) $ where the effective Hamiltonian
\begin{equation}
H_{\mathrm{eff}}=H_{A}+H_{S}+\frac{1}{2}\left[ V_{AS},S\right] ,
\end{equation}%
is just the Frohlich-Nakajima Hamiltonian in solid state physics \cite%
{frohlich}. Here, we re-derive it to justify its applicability in
thermodynamics.

Since the minimal energy level spacing of the system is much larger than the
energy level spacing of the apparatus, the effective interaction $V_{\mathrm{%
eff}}$ can be obtained as $V_{\mathrm{eff}}=\sum_{n}\mathcal{H}\left(
n\right) \left\vert n\right\rangle \left\langle n\right\vert $ where
\begin{equation}
\mathcal{H}\left( n\right) =\sum_{k}G(n)\left\vert k\right\rangle
\left\langle k^{\prime }\right\vert
\end{equation}%
is a branched effective Hamiltonian of the apparatus corresponding
to the situation that the system is prepared in the state
$\left\vert n\right\rangle $ with the coupling
\begin{equation}
G(n)=\frac{\left\langle nk\right\vert V_{AS}\left\vert nk^{\prime
}\right\rangle }{\left( \epsilon _{k}-\epsilon _{k^{\prime }}\right)}.%
\end{equation}
The above obtained effective interaction $V_{\mathrm{eff}}$
satisfies $\left[
H_{S},V_{\mathrm{eff}}\right] =0$. Thus the total effective Hamiltonian $H_{%
\mathrm{eff}}=H_{S}+H_{A}+V_{\mathrm{eff}}$ describes a
non-demolition measurement without the presence of an environment.
In this case, the factorized initial state $\left\vert \varphi
\left( 0\right) \right\rangle =\sum_{n}c_{n}\left\vert
n\right\rangle \otimes \left\vert D\right\rangle $ will evolve into
$\left\vert \varphi \left( t\right) \right\rangle
=\sum_{n}c_{n}\left\vert n\right\rangle \otimes \left\vert
D_{n}\left( t\right) \right\rangle $ where $\left\vert D_{n}\left(
t\right) \right\rangle $ takes the form $\left\vert D_{n}\left(
t\right) \right\rangle =\exp \left[ -iG\left( n\right) t\right]
\left\vert D\right\rangle .$ If $\left\{ \left\vert D_{n}\left(
t\right) \right\rangle \right\} $ becomes an orthogonal set when $t$
approaches infinity, then the time evolution $\left\vert \varphi
\left( t\right) \right\rangle $ represents a process of ideal
pre-measurement.

\textit{Thermodynamic effects of measurements}- Now we study the change in
the thermodynamic features of the system caused by the apparatus. We assume
that both the probed system and the measuring apparatus are immersed in the
same thermal bath with temperature $T$ or inverse temperature $\beta
=1/k_{B}T$. After or during the measuring process, the total system $S+A$
will reach the state with the same temperature $T$, if the total system is
non-degenerate. Then we can calculate the reduced density matrix $\rho _{S}=%
\mathrm{Tr_{A}}\left[ \exp \left( -\beta \left( H_{S}+H_{A}+V_{\mathrm{eff}%
}\right) \right) \right] $, obtaining
\begin{equation}
\rho _{S}=\frac{1}{Z_{S}^{\prime }}\sum_{n}e^{-\beta E_{n}}\xi
\left( n\right) \left\vert n\right\rangle \left\langle n\right\vert
,  \label{eq:rho}
\end{equation}%
where $Z_{S}^{\prime }=\sum_{n}\exp \left( -\beta E_{n}\right) \xi \left(
n\right) $ and
\begin{equation}
\xi \left( n\right) =\mathrm{Tr_{A}}\left\{ \left\langle n\right\vert \exp %
\left[ -\beta \left( H_{A}+V_{\mathrm{eff}}\right) \right] \left\vert
n\right\rangle \right\}
\end{equation}%
is a formal factor depending on the system state $\left\vert n\right\rangle $
and vanishing trivially when no coupling exists.

\begin{figure}[tbp]
\includegraphics[bb=40 407 515 724, width=6cm, clip]{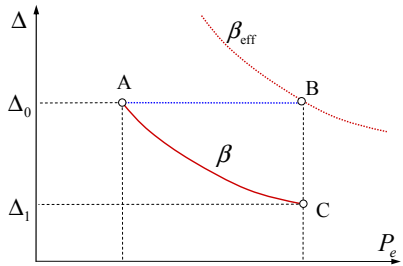}
\caption{Two interpretations. (i) Isometric process ($A\rightarrow
C$, the solid line): the inverse temperature $\protect\beta$ is
fixed in the process. (ii) Isothermal process ($A\rightarrow B$, the
dotted line): The level spacing $\Delta$ is fixed in the process.}
\label{fig:2pro}
\end{figure}

Here, we consider a manipulation process. (i) Initially, no probing
apparatus is coupled to the system, which is in equilibrium with the
heat bath with temperature $T$. (ii) Then, the apparatus begin to
probe the system at time $t=0$. As the evolution described in
Fig.~\ref{fig:2pro}, the total system $S+A$ reaches the state with
the same temperature $T$. We observe that the thermodynamic effect
of measurement implied in Eq.~(6) allows two interpretations. These
two interpretations are illustrated in Fig.~\ref{fig:2pro} as those
in Ref.~\cite{q1,quanPRE2007}.

The first interpretation goes as follows. The change from the close thermal
state of the system $\rho_{Sc}=\exp\left(-\beta H_{S}\right)/Z_{S}$ to the
modified thermal state $\rho_{S}$ is understood as a quantum isometric
process, represented by the solid line between the points $A$ and $C$ in Fig.~%
\ref{fig:2pro}. In this process, the temperature keeps unchanged, but the
energy level spacings are alternated by the interaction with the apparatus.
Accordingly, we define $\xi\left(n\right)=\exp\left(-\beta\Delta
E_{n}\right) $, then $\rho_{S}$ can be written in the form with clear
physical meaning:
\begin{equation}
\rho_{S}=\frac{1}{Z^{\prime }_{S}}\sum_{n}e^{-\beta\left(E_{n}+\Delta
E_{n}\right)}\left|n\right\rangle \left\langle n\right|.
\end{equation}
In the present case, it is rather natural to regard the inner energy change $%
\Delta U=\mathrm{Tr_{S}}\left[H_{\mathrm{S}}\left(\rho_{S}-\rho_{Sc}\right)%
\right]$ as a witness of the thermodynamic role of measurement.

In the second interpretation the change from $\rho _{Sc}$ to $\rho
_{S}$ is understood as a quantum isothermal process, represented by
the dotted line between the points $A$ and $B$ in
Fig.~\ref{fig:2pro}. In this process, the energy level spacings are
fixed. To justify our using the term \textquotedblleft isothermal
process" here, at least to some extent, we define the effective
temperature \cite{efftem} via $\beta \left( n\right) =\ln
(P_{n}/P_{n+1})/\Delta _{n}$ or
\begin{equation}
\beta \left( n\right) =\beta +\frac{1}{\Delta _{n}}\ln \frac{\xi \left(
n\right) }{\xi \left( n+1\right) },
\end{equation}%
where $\Delta _{n}=E_{n+1}-E_{n}$ is the $n-th$ energy level spacing. We
notice that generally this generalized temperature cannot be regarded as an
effective temperature since it depends on the energy levels of the system.
But if $\beta \left( n+1\right) -\beta \left( n\right) =0,$ i.e.,
\begin{equation}
\left[ \frac{\xi \left( n+1\right) }{\xi \left( n+2\right) }\right] ^{\Delta
_{n}}=\left[ \frac{\xi \left( n\right) }{\xi \left( n+1\right) }\right]
^{\Delta _{n+1}},
\end{equation}%
then $\beta \left( n\right) $ becomes a well defined thermodynamic parameter
independent of $n$, which we denote by $\beta _{\mathrm{eff}}$. In this
case, the above defined generalized temperature allows the physical
interpretation of effective temperature, and we have $\rho _{S}=\sum_{n}\exp
\left( -\beta _{\mathrm{eff}}E_{n}\right) \left\vert n\right\rangle
\left\langle n\right\vert /Z_{S}^{\prime }$, and the inner energy change
\begin{equation}
\Delta U=\sum_{n}E_{n}\left[ e^{-\beta _{\mathrm{eff}}E_{n}}-e^{-\beta E_{n}}%
\right]
\end{equation}%
reflects the thermalization effect.

\textit{The quantum role of measurement and renormalization}- As an
explicit illustration, we model the apparatus with weak couplings to
the probed system as a collection of harmonic oscillators. According
to the results of Ref.~\cite{leggett} and Ref.~\cite{suncp98pra}
(where he arguments are carried out for the bath modeling, but can
work well for our setup for the apparatus), the coupling of the
system to the apparatus  is linear with respect to the coordinates
of the bath harmonic oscillators. Let $b_{j}^{\dagger }\left(
b_{j}\right) $ be the creation(annihilation) operator of the $j-th$
mode of the bath with eigen-frequency $\omega _{j}$,
and $\lambda _{n}g_{j}$ the coupling coefficient of the system state $%
\left\vert n\right\rangle $ to the $j-th$ mode. Then, the total Hamiltonian
is obtained as
\begin{eqnarray}
H &=&\sum_{n}E_{n}\left\vert n\right\rangle \left\langle n\right\vert
+\sum_{k}\omega _{k}b_{k}^{\dagger }b_{k}  \nonumber \\
&&+\sum_{n}\lambda _{n}\left\vert n\right\rangle \left\langle n\right\vert
\sum_{k}\left( g_{k}a_{k}^{\dagger }+\mathrm{h.c}\right) .
\end{eqnarray}%
In this case, the above defined generalized inverse temperature reads
\begin{equation}
\beta \left( n\right) =\beta \left( 1-\frac{|\lambda _{n+1}|^{2}-|\lambda
_{n}|^{2}}{E_{n+1}-E_{n}}\varepsilon \right) ,
\end{equation}%
where
\begin{equation}
\varepsilon =\sum_{k}\frac{|g_{k}|^{2}}{\omega _{k}}=\int \rho \left( \omega
\right) \frac{|g_{k}|^{2}}{\omega _{k}}dk
\end{equation}%
represents the self-energy of the apparatus, which causes the Lamb shift of
its coupled system. As pointed above, to define reasonably an effective
temperature for the system, the generalized inverse temperature $\beta
\left( n\right) $ should be independent of the energy level. Here is a
simple example satisfying this condition: the system is a harmonic
oscillator with the energy level $E_{n}=\left( n+1/2\right) \omega $ and the
coupling strength $\lambda _{n}=\sqrt{n}$. In this example, the well defined
effective inverse temperature is
\begin{equation}
\beta _{\mathrm{eff}}=\beta (1-\sum_{k}\frac{|g_{k}|^{2}}{\omega \omega _{k}}%
),
\end{equation}%
and the corresponding effective temperature $T_{\mathrm{eff}}=\left(
k_{B}\beta _{\mathrm{eff}}\right) ^{-1}$ of the system is higher than that
of the environment $T=\left( k_{B}\beta \right) ^{-1}$.

For a two-level system with the excited state $\left\vert e\right\rangle $
and the ground state $\left\vert g\right\rangle $ and the energy level
spacing $\Delta $, the effective inverse temperature is also well defined.
It reads
\begin{equation}
\beta _{\mathrm{eff}}=\beta +\frac{1}{\Delta }\ln \frac{\xi _{g}}{\xi _{e}}.
\end{equation}%
If the apparatus is a single mode cavity with frequency $\omega
_{b}$ and the two-level system is coupled to it by the dipole
interaction with coupling strength $g$, then in the large detuning
case $\omega _{b}\gg \Delta $, the formal factor $\xi _{e}$ and $\xi
_{g}$ can be explicitly calculated as follows:
\begin{eqnarray}
\xi _{e} &=&\sum_{n}\exp \left\{ -\beta \left[ \omega _{b}n+\frac{\left\vert
g\right\vert ^{2}}{\omega _{b}-\Delta }\left( n+1\right) \right] \right\} ,
\nonumber \\
\xi _{g} &=&\sum_{n}\exp \left\{ -\beta \left[ \omega _{b}n+\frac{\left\vert
g\right\vert ^{2}}{\omega _{b}-\Delta }n\right] \right\} .
\end{eqnarray}

Since $\xi_{g}>\xi_{e}$, we are led to the conclusion that the measurement
will decrease the temperature of the system by
\begin{equation}
\Delta T=\frac{\frac{1}{\Delta}\ln\frac{\xi_{g}}{\xi_{e}}}{\beta+\frac{1}{%
\Delta}\ln\frac{\xi_{g}}{\xi_{e}}.}T.
\end{equation}

\begin{figure}[tpb]
\includegraphics[width=7cm, bb=49 102 271 372]{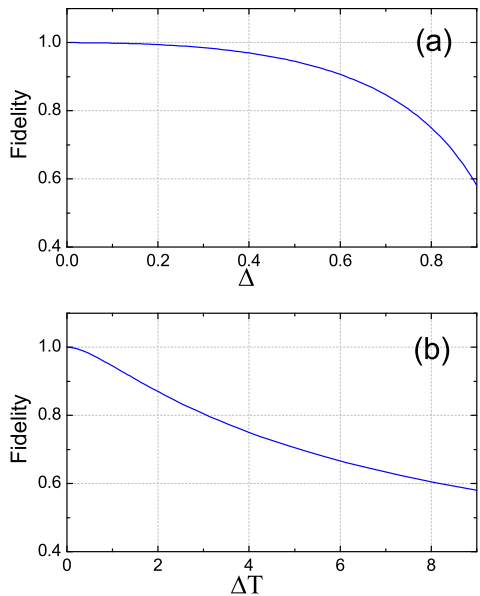}
\caption{Fidelity. (a) Fidelity $\mathcal{F}$ vs the level shift
$\lambda$.
(b)Fidelity $\mathcal{F}$ vs the temperature shift $\Delta T$. Here, we take $%
\protect\omega=1$ and $\protect\beta=1$.} \label{Fig:fedality}
\end{figure}

Finally, to quantitatively evaluate the thermalized state resulting
from the interaction with the environment, let us check the fidelity
\cite{fed} $\mathcal{F}$ of such a state to the ordinary canonical
state $\rho _{\mathrm{can}}=\exp \left( -\beta H_{S}\right)
/\mathrm{Tr}\left[ \exp \left( -\beta H_{S}\right) \right] $.
Generally for  the reduced density matrix in Eq.~\ref{eq:rho}, the
fidelity between the initial state and the final state reads
\begin{equation}
\mathcal{F}=\frac{\sum_{n}e^{-\beta E_{n}}\sqrt{\xi \left( n\right) }}{%
\left( \sum_{n}e^{-\beta E_{n}}\right) ^{1/2}\left( \sum_{n}e^{-\beta
E_{n}}\xi \left( n\right) \right) ^{1/2}}.
\end{equation}%
When the probed system is also a harmonic oscillator discussed above, the
fidelity can be analytically obtained as
\begin{equation}
\mathcal{F}=\frac{\sqrt{\sinh \frac{\beta \omega }{2}\sinh \frac{\beta
\left( \omega -\lambda \right) }{2}}}{\sinh \frac{\omega -\lambda /2}{2}},
\end{equation}%
where $\lambda =\sum_{k}|g_{k}|^{2}/\omega _{k}$ characterizes the
shift of the energy level of the harmonic oscillator, which reflects
the  effect due to the coupling to the apparatus. In
Fig.~\ref{Fig:fedality}(a), the fidelity is plotted against the
level shift. As the coupling between the apparatus and the system is
turn on, the energy level for the system is effective
$E'_n=n(\omega-\lambda)$ based on the first interpreting of the
reduced density matrix. The fidelity between the canonical thermal
and the reduced density matrix decreases as the coupling becomes
strong. Therefore, the system gradually deviates from canonical
thermal state, leaving an evidence of witness of the apparatus. For
the second interpreting, we plot the fidelity $\mathcal{F}$ as the
function of the temperature shift $\Delta
T=1/\beta_{\mathrm{eff}}-1/\beta=\left [
\beta(\omega/\lambda-1)\right]^{-1}$ in Fig.~\ref{Fig:fedality}(b).

\textit{Conclusion}- In summary, for the weak coupling case, we show from
the generalized approach of Frohlich-Nakajima transformation the
universality of a non-demolition Hamiltonian in connection with the probing
process. Based on this general non-demolition measurement, we investigate
the probing effect on the system when the total measurement happens in a
reservoir. It is concluded that the probing of the system can be witted
through the change of the effective temperature, even though there is no
direct energy exchange between the system and the detector. To characterize
the change of state for the system, we evaluate the fidelity of the modified
canonical thermal state to the original canonical state without being probed.

The work is supported by National Natural Science Foundation of China and
the National Fundamental Research Programs of China under Grant No. 10874091
and No. 2006CB921205.

\end{document}